\documentclass[10pt,letterpaper]{article}
\usepackage{opex3}
\pdfoutput=1

\newcommand{\vecb}[1]{\mbox{\boldmath$#1$}}

\newcommand{\um}{$\mu$m}

\usepackage{amssymb}
\usepackage{amsmath}
\usepackage{latexsym}

\newcommand{\figFingerprintingSchematics}{aom_schematics_lens_2_copy_nb.pdf}

\newcommand{\figFingerprintingTwoP}{figureFingerprExpt.pdf}
\newcommand{\figBragg}{bragg_imagingBlue.pdf}

\begin{document}

\title{Super-resolution imaging via spatiotemporal frequency shifting and
  coherent detection}

\author{Leonid Alekseyev$^1$, Evgenii Narimanov$^1$, and Jacob Khurgin$^2$}
\address{${}^1$ Department of Electrical and Computer Engineering,\\ Purdue University, West Lafayette, IN 47907}
\address{${}^2$ Department of Electrical and Computer Engineering,\\ Johns Hopkins University, Baltimore, MD 21218}

\email{evgenii@purdue.edu}

\begin{abstract} \noindent 
  Diffraction limit is manifested in the loss of high spatial frequency
  information that results from decay of evanescent waves.  As a result,
  conventional far-field optics yields no information about an object's
  subwavelength features.  Here we propose a novel approach to recovering
  evanescent waves in the far field, thereby enabling subwavelength-resolved
  imaging and spatial spectroscopy.  Our approach relies on shifting the
  frequency and the wave vector of near-field components via scattering on
  acoustic phonons.  This process effectively removes the spatial frequency
  cut-off for unambiguous far field detection.  This technique can be adapted
  for digital holography, making it possible to perform phase-sensitive subwavelength
  imaging.  We discuss the implementation of such a system in the mid-IR and
  THz bands, with possible extension to other spectral regions.
\end{abstract}



\section{Introduction}

Microscopic imaging is the oldest and one of the most important non-invasive
analysis techniques.  It has been immensely successful in uncovering the
structure, composition, and dynamics of micro- and nanoscale chemical and
biological samples.  Contemporary investigations in the life sciences demand
ever-increasing resolution in a variety of spectral bands, with much
attention given to IR and THz.  This task is made complicated by the
diffraction limit, which sets a fundamental upper bound on the maximum
spatial frequency conveyed by conventional refractive optics.

Between the initial studies of optical resolution by Rayleigh and Abbe in the
19th century and the present day, a multitude of super-resolution systems and
methods have been proposed and demonstrated~\cite{Testorf2010}.  Extracting
information beyond the diffraction limit remains an active area of research.
One general strategy for achieving sub-diffraction-limited images originates
in the fact that it is possible to increase the spatial bandwidth of an
optical system by sacrificing certain other characteristics (e.g.\ field of
view, temporal bandwidth, acquisition time).  This idea was pioneered by
Lukosz in his seminal 1967 paper~\cite{Lukosz1967}.  Indeed, many
contemporary super-resolution techniques can be viewed as implementations of
the Lukosz approach~\cite{Testorf2010,Lukosz1967}, including off-axis
illumination~\cite{Kuznetsova2007}, structured
illumination~\cite{Gustafsson2005}, spatial frequency-shifting
gratings~\cite{Paturzo2008,Durant2006}, and even near-field scanning
microscopy~\cite{Dragnea2001,Keilmann1999,Planken2002}.

The idea of improving spatial resolution by using temporal degrees of freedom
is particularly appealing for two reasons. First, time multiplexing can help
resolve ambiguities that arise when high spatial frequencies are scattered
into the optical passband, as in the case of evanescent waves diffracting off
a subwavelength grating~\cite{Durant2006}.  Second, in certain frequency
bands outside the visible spectrum (e.g.\ far-IR or THz) it might be easier
to manipulate signals in the time domain than in the spatial frequency
domain.  The potential utility of time-domain (or temporal-frequency domain)
for super-resolution has long been recognized; in fact, a
frequency-multiplexing scheme involving conjugate moving gratings was
described in Lukosz's original paper~\cite{Lukosz1967} and demonstrated
several decades later~\cite{Mendlovic1997a,Shemer1999}.  However, owing to
the complexity of the experimental setup, this scheme has not seen wide
adoption.

In the present paper, we propose a time-multiplexed super-resolution system
that requires no moving parts and is based on coherent detection of a
frequency-shifted signal.  This scheme lends itself particularly well to
super-resolved imaging in IR and THz.  Our approach is based on a device that
converts evanescent waves to propagating waves via diffraction on acoustic
phonons.  The scattered and frequency-shifted waves can be easily decoupled
from the existing propagating spectrum that forms the regular
diffraction-limited image (thus, the image is free from aliasing).  With
minimal processing, these shifted components can be used to distinguish
subwavelength features.

We will discuss two variations of this approach.  Both rely on mixing the
frequency-shifted fields scattered from the object with a reference wave,
creating, at the detector, a beat note photocurrent which can be isolated
through lock-in techniques.  We will see that detection of high spatial
frequencies is enabled by the spatial frequency offset of the scattered
signal, and that a true super-resolved imaging configuration can be attained
with an additional temporal frequency shift of the reference signal.

\section{Scattering from a phonon grating: general description}
\begin{figure}
\centerline{\scalebox{.48}{\includegraphics{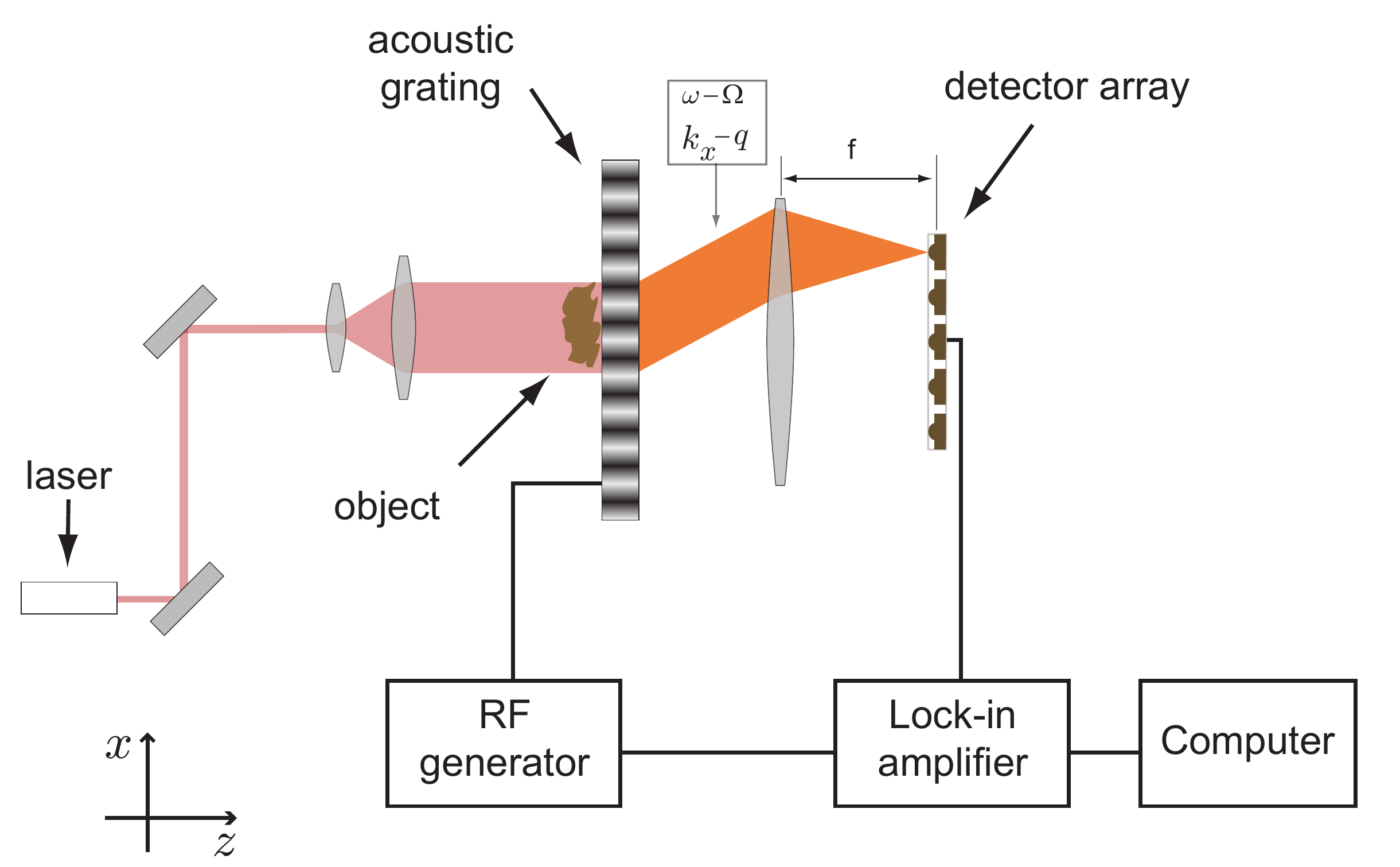}}}
\caption{Schematics of the proposed
  system. } \label{fig:fingerprintingSch}
\end{figure}

The proposed super-resolution microscope/sensing system is
shown in Fig.~\ref{fig:fingerprintingSch}. The object is placed in the
near field of an acousto-optic modulator (AOM) and illuminated with a
plane wave from a mid-IR or THz source.  Waves scattered from the
object strike the phonon grating set up in the AOM by a running
acoustic wave with frequency~$\Omega$.  Due to scattering on the
phonons, the transverse wave vector $k_x$ of the incident radiation is
shifted by integer multiples of the phonon wave vector $q$, while its
corresponding frequency is shifted by integer multiples of~$\Omega$.
For a sufficiently large $q$, the evanescent components of the
object's spatial spectrum ($|k_x| > \omega/c$) can be scattered into
the propagating waves with $|k'_x| \equiv |k_x - q| < \omega/c$.  The
various spatial frequency components can be measured using a Fourier
optics setup (e.g.\ a lens with a detector array in its focal plane).

We model our system by considering a rectangular sound column (i.e.\ planar
acoustic wavefronts propagating in the $x$ direction) interacting with a
spectrum of incident plane waves in a dielectric medium.  We neglect the
diffraction of the sound field or multiple reflections, and assume weak
interaction.  Due to photoelastic effect, the sound field produces a
sinusoidal modulation of the dielectric permittivity~\cite{Boyd_NLO},

\begin{equation}\label{eq:epsilon_wave}
\epsilon(x) = \overline{\epsilon} + \Delta \epsilon \cos(q x-\Omega
t),\end{equation} which corresponds to a spatiotemporal volume
grating.  We may write the general form of the field inside the
grating as a sum over the discrete diffracted orders,
\begin{equation}\label{eq:scatt_sum}
E = \sum_j A_j(z) \exp[i(k_x + jq)x - i(\omega+j\Omega)t].
\end{equation}  The scattered plane wave components $A_j(z)$ are then governed by the Raman-Nath equations~\cite{KorpelBook},

\begin{equation}\label{eq:recursion}
A''_j(z)+k_{z_j}^2 A_j(z) = -\frac{\Delta \epsilon}{2 c^2} \omega_j^2
[A_{j-1}(z) + A_{j+1}(z)],
\end{equation}
where $k_{z_j} = \left[\overline{\epsilon}\frac{\omega^2}{c^2} - (k_x
  + j q)^2\right]^{1/2}$, and $\omega_j \simeq \omega$.

The amplitude of the $j^{\rm th}$ diffracted order, $A_j$, is
proportional to $(\Delta\epsilon)^j$, with $\Delta\epsilon/\epsilon
\ll 1$, allowing to ignore higher order terms ($j \ge 2$).  We can,
furthermore, conclude that the amount of energy scattered into the
shifted waves is small, thereby permitting to neglect the variation of
$0^{\rm th}$ diffracted order $A_0$ (the undepleted pump
approximation)~\cite{Boyd_NLO}. We note that for propagating waves,
this conclusion is valid insofar as there exists no Bragg matching
between the incident and diffracted waves.  Since the phonon wave
vector $q$ is a tunable parameter in our model, it is always possible
to pick a range of $q$ values to ensure minimal energy loss in the
incident wave. For the evanescent waves, the undepleted pump
approximation is justified by the small interaction length.

Keeping terms up to first order in Eq.~(\ref{eq:scatt_sum}), we
obtain:

\begin{equation}
A''_\pm(z)+k_{z_j}^2 A_\pm(z) = -\frac{\Delta \epsilon}{2 c^2}
\omega_j^2 A_0(z).\end{equation} The scattering amplitudes of
``upshifted'' and ``downshifted'' waves can be obtained from this
expression.  Since the input field at spatial frequencies $(k_x \mp
q)$ contributes to the output field at spatial frequency $k_x$, we may
write:
\begin{equation}\label{eq:spec_out}
E_{\rm out}(k_x) = \left[\tilde{A}_- \exp(i\Omega t) + \tilde{A}_+
  \exp(-i\Omega t) + \tilde{A}_0\right]\exp[i(k_x x - \omega
  t)],\end{equation} with $\tilde{A}_\pm \equiv t^\pm E_{\rm in}(k_x
\mp q)$, $\tilde{A}_0 \equiv t_0 E_{\rm in}(k_x)$. Scattering coefficients $t^\pm$
are, for the case of evanescent waves, given by

\begin{equation}\label{eq:tpm} t^\pm = t_0
\frac{\Delta\epsilon}{2}\left(\frac{\omega}{c}\right)^2
\frac{1}{k_z^\pm\sqrt{(k_z^\pm)^2 + \kappa^2}}
\approx
t_0
\frac{\Delta\epsilon}{2}\left(\frac{\omega}{c}\right)^2
\frac{1}{k_z^\pm\sqrt{q(q\mp2k_x)}},
\end{equation} where 
$k_z^\pm = \sqrt{\epsilon(\omega/c)^2-(k_x\mp q)^2}$, $\kappa = i k_z =
\sqrt{k_x^2-\epsilon(\omega/c)^2}$, and $t_0$ is a Fresnel transmission
coefficient.  We see that the conversion of incident evanescent waves into
propagating signals depends critically on the acoustooptic index contrast
$\Delta\epsilon$ and on the effective interaction length $1/\kappa$.
We note, also, that Eq.~\eqref{eq:tpm} describes also the generation of
shifted spatial
frequencies for the case where the incident wave is propagating (provided we
make the association $\kappa=i k_z$).  In this case, the divergence of
Eq.~\eqref{eq:tpm} around 
$q\approx 0$, as well as $q\approx \pm2
k_x$ signifies the breakdown of the perturbative treatment of Eq.~(\ref{eq:recursion}) 
due to the onset of Bragg-matching.

From Eq.~(\ref{eq:tpm}) we can estimate the diffraction efficiency of
high spatial frequency input signal $E_{\rm in}(k_x^{\rm in})$ as
\begin{equation}\label{eq:diffeff}
\left|t^\pm\right| \approx \frac{\omega/c}{2k_x^{\rm in}}
\frac{\Delta\epsilon}{n(1+n)},
\end{equation} with 
$n=\sqrt{\epsilon}$ (the refractive index of the acoustic medium), and
$\Delta\epsilon\propto\sqrt{F}$, the flux of acoustic energy per unit
area.

We thus obtain the amplitudes of the frequency-shifted waves.  We define $\tilde{A}_\pm = t^\pm E_{\rm in}(k_x \mp q)$,
$\tilde{A}_0 = t_0 E_{\rm in}(k_x)$ (with linear coefficients $t_0$,
$t^\pm$ describing the generation of phonon-scattered and/or device
transmission characteristics) and assume $\tilde{A}_i \gg \tilde{A}_0
\gg \tilde{A}^\pm$, where $\tilde{A}_i$ is the detected amplitude of
the illuminating wave $A_i e^{i k_0 z}$.  Averaging out the signal
over the finite detector aperture and subtracting the background
(which can be done electronically), we can write the intensity
detected by the system of Fig.~\ref{fig:fingerprintingSch} as

\begin{equation}\label{eq:measurement}
I_{\rm out} = |\tilde{A}_0|^2 + 2
\left(|\tilde{A}_i\tilde{A}_-|^2+|\tilde{A}_i\tilde{A}_+|^2\right)^{1/2}
\cos(\Omega t+ \gamma).
\end{equation}
The two terms in this equation can be decoupled using standard
techniques: the DC term is isolated with the aid of a low-pass filter,
while the term oscillating at the acoustic frequency $\Omega$ is
recoverable using standard lock-in detection.  For any given $k_x$, this second
term contains contributions from both $\tilde{A}_+ = t^+ E_{\rm
  in}(k_x-q)$ and $\tilde{A}_-=t^- E_{\rm in}(k_x+q)$. Although the
coupling between these two quantities, together with the lack of phase
information, makes it difficult to recover the spatial spectrum, the
information collected can be used in detecting subwavelength
morphological changes between different samples.

\section{Super-resolved fingerprinting: numerical simulation}

We now illustrate the ability of the proposed system to distinguished between
subwavelength spatial features of different objects. In particular, we
utilize Eq.~(\ref{eq:measurement}) to perform a comparison between the
standard optical target (USAF test chart) and a modified target,
where the label of every 6th line group has been randomly replaced.  The
first replacement corresponds to the last resolvable line group
($\lambda$/2.5 line separation); the subsequent replacements correspond to
halving the size of the line groups ($\lambda$/5,$\ldots$, $\lambda$/40).  We
assume the measurement is performed by selecting an element of a
photodetector array in the observation plane and using two orthogonal
acoustic transducers to scan the acoustic wavevector within the range
$q_{x,y} \in [-25\,\omega/c, 25\,\omega/c]$.  (We choose these values with
the aim to improve by a factor of $\sim20$ on the Abbe $\lambda/2$ resolution
limit, thereby collecting meaningful information about the $\lambda/40$ line
group.) 

In our computations, we assume the operating wavelength of 10
\um\ with germanium as the acoustic medium.  We take
$\Delta\epsilon=10^{-3}$ and restrict the magnitude of the acoustic
wave vector $q$ to $25\,\omega/c$.  Since for high
spatial frequencies $k_x^{\rm in} \approx q$, acoustic driving
frequencies up to 8.75 GHz are required to retrieve $k_x^{\rm in}
\approx 25 \omega/c$.  These parameters are within reach of modern
ultrasonic transducers~\cite{HiFreqUltrasoundBulk}, as well as surface
acoustic wave devices~\cite{HiFreqUltrasoundSAW}. 

\begin{figure}
\centerline{\scalebox{.35}{\includegraphics{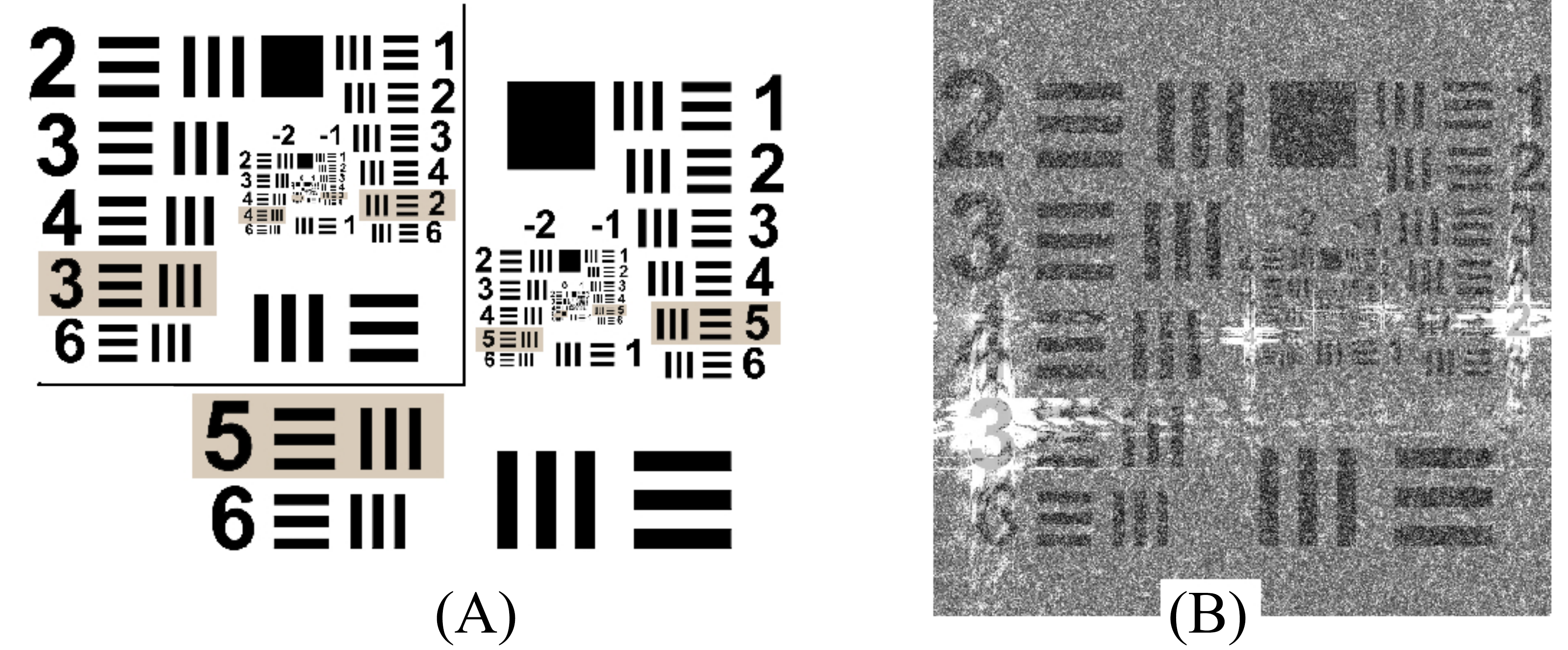}}}
\caption{(a) Optical test target and its modified version (inset).  In
  the modified target, the ``5'' label of every column has been
  replaced by another digit.  (b) Computed output of the system in the
  presence of noise (shown in grayscale) assuming a realistic, noisy
  detector with 400 active photocells.  The modified optical target is
  superimposed for illustration purposes.  The output of the system
  clearly identifies the location of every modified digit, even for
  regions far below the diffraction limit.} \label{fig:fingerprinting}
\end{figure}

It should be emphasized that any method that relies on digital
processing of raw data can suffer form rapid -- sometimes
exponential~\cite{PodolskiyNarimanovNSSL} -- accumulation of noise.
(Indeed, some proposed super-resolution methods even stipulate the need for an
exponentially strong input signal to overcome noise and losses~\cite{Zheludev2008a}.)
To show that this is not the case here, in our computations we add a
normally-distributed random term to the AC amplitude of
Eq.~(\ref{eq:measurement}) in order to simulate noise in the system.
Because SNR is expected to be lowest for maximum values of the
acoustic wavevector $q$, we consider SNR=10 for
$q=25\,\omega/c$~\footnote{Acoustooptic diffraction efficiency, and
  hence the signal-to-noise ratio varies as $1/q$.}.  Assuming a
 20$\times$20 element photodetector array, we compute the
signal given by Eq.~(\ref{eq:measurement}) for the standard target, as
well as the modified target [Fig.~\ref{fig:fingerprinting}(a)].
Fig.~\ref{fig:fingerprinting}(b) shows the result of subtracting the
two datasets and performing an inverse Fourier transform, with the
resulting plot superimposed onto the modified optical target.
Evidently, every change in the original image is manifested in this
difference diagram.  Furthermore, it is largely localized in the
vicinity of the actual changed pixels.  It is possible to discern the
difference signal even from the $\lambda$/40 line group label.

The ability to distinguish between fine spatial features of optical
targets makes the system described above uniquely suited for
identifying objects based on their subwavelength spatial features.  As
a result, it may find applications in fingerprinting and/or
detection of chemical and biological structures.

\section{Super-resolved digital holography}
A straightforward modification of the setup described above not only allows
to measure the ``downshifted'' $\tilde{A}_-$ component directly, but also
provides a method for retrieving phase information, making it possible to
perform phase-contrast microscopy, and potentially enabling 3D imaging on
subwavelength scales.

  To this end, a portion of the illuminating radiation is shifted in
  frequency by $\Omega_b$ using a second AOM.  Unlike the modulator
  that interacts with light scattered from the sample in the
  Raman-Nath regime~\cite{Boyd_NLO}, this second AOM utilizes an
  appropriately oriented and longer cell to produce Bragg scattering.
  This results in a strong optical signal at frequency
  $\omega+\Omega_b$, $|\tilde{A}_b|\exp[i(k_b\cdot r - (\omega +
    \Omega_b)t)]$, which is projected onto the detector [see
    Fig.~\ref{fig:bragg}(a)].  Interference between the two optical
  signals produces beat note photocurrents with frequencies $\Omega$,
  $\Omega_b$, $\Omega_b+\Omega$, $\Omega_b-\Omega$:

\begin{figure}
\centerline{\scalebox{.32}{\includegraphics{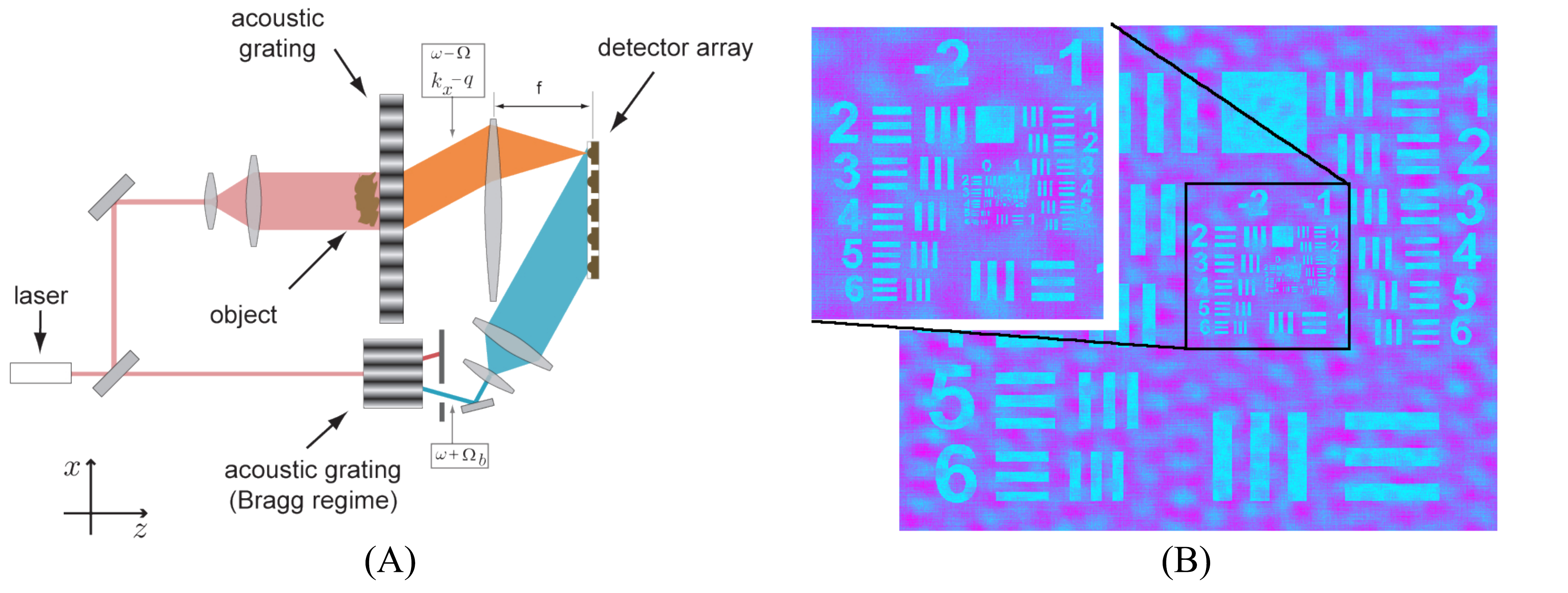}}}
\caption{(a) Schematics of the proposed system. Note the Bragg-shifted
  reference beam that aids in providing phase information.  (b) The
  computed output of the system with optical test target as the object
  in the presence of noise.} \label{fig:bragg}
\end{figure}

\begin{align}\label{eq:iOut}
I_{\rm out}(k_x) = {} & \left|E_i \exp(i k_0 z) + 
  \tilde{A}_b\exp(i\vecb{k}\cdot\vecb{r})
  + [\tilde{A}^- \exp(i\Omega t) +
  \tilde{A}^+ \exp(-i\Omega t) + 
  \tilde{A}_0]\exp(i\vecb{k}\cdot\vecb{r})\right|^2 \nonumber \\ 
= {} & \ldots  + 2 |\tilde{A}^-\tilde{A}_b|
\cos[(\Omega_b+\Omega)t+\Delta\Phi^-] 
  + 2 |\tilde{A}^+\tilde{A}_b|
 \cos[(\Omega_b-\Omega)t+\Delta\Phi^+]+\ldots,
\end{align}
where $\Delta\Phi^\pm = (k_b-k)\cdot r - \phi^\pm$ is the phase
difference between the signal from the Bragg cell,
$|\tilde{A}_b|\exp(i k_b\cdot r)$, and the Raman-Nath-scattered signal
$\tilde{A}^\pm\exp(i k\cdot r) = |\tilde{A}^\pm|\exp[i(\phi^\pm +
  k\cdot r)]$.

Of special interest is the component at frequency $\Omega+\Omega_b$,
which carries the high spatial frequency information contained in its
modulus and its phase $\Delta\Phi^-\simeq (k_x^b-k_x)x - \phi^-$.
Both of these quantities can be retrieved using lock-in techniques.
To produce the lock-in reference, the RF signals driving the two
acoustic cells can be mixed using a nonlinear element (e.g.\ a diode)
and appropriately filtered to produce the sum frequency.  As a result,
complete information can be obtained about the complex high spatial
frequency Fourier component $\tilde{A}^-$, from which it is
straightforward to deduce the field $E_\text{in}(k_x+q)$. By
collecting data from multiple CCD pixels, as well as by varying the
acoustic wave vector $q$, information can be collected about the
entire spatial spectrum of the object.  The data can then be digitally
processed to produce a spatial-domain image containing subwavelength
details, as well as phase contrast.

Because the Bragg-shifted signal we use to decouple the $\tilde{A}_+$ and
$\tilde{A}_-$ terms serves as a reference needed to record phase information,
and because the image is reconstructed digitally, our technique belongs in
the category of digital Fourier holography
(DFH)~\cite{OriginalGoodmanDigitalHoloPaper,HoloCCDProblem1,CucheDigitalHolo}.
However, the proposed coherent detection method is different from traditional
DFH setups in that it effectively converts sample spatial frequencies to
temporal ones.  In conventional holography, care has to be taken to isolate
the target signal both in real and Fourier space.  This translates into
limitation on the field of view, as well as maximum attainable
resolution~\cite{CucheDigitalHolo}.  The requirement that the CCD pixel
spacing must allow for imaging the reference wave fringes further limits the
resolution.  As a result, simple digital holography setups suffer from very
narrow spatial frequency passbands (N.A.$\sim$0.1).  Synthetic aperture
techniques based on gratings~\cite{Alexandrov2006,Liu2002,Paturzo2008} have
been successfully shown to increase the effective numerical aperture, as has
heterodyne detection~\cite{LeClerc2000}.  Our method combines the benefits of
these approaches: dynamic acoustic grating allows to scan the spatial
frequency space while always remaining in the CCD's passband, while the
temporal frequency shifts multiplex the data, effectively increasing the
number of information channels in a given spatial frequency band.

We simulate the performance of the frequency-shifted digital holography
system by first using Eq.~(\ref{eq:iOut}) to compute the response of the
system to a calibration signal having unit amplitude for all spatial
frequencies.  In practice, such calibration signal might be generated by
placing a point source in the vicinity of the AOM.  Eq.~(\ref{eq:iOut}) also
provides the effective amplitude and phase transfer functions that allow to
determine the detected signal for a given input field distribution.  Gaussian
noise is added to simulate spurious signals in the system.  The input signal
can then be obtained by dividing out the calibration quantities.  In
Fig.~\ref{fig:bragg}(b) we plot the simulated retrieved field magnitude.
Zooming in on the central part of the test patten (figure inset) it is
evident that every line group is distinctly resolved, suggesting that the
effective numerical aperture is $>1$ (due to reconstruction of evanescent
waves).  

\section{Potential improvements of the proposed system and extension to higher frequencies}
As demonstrated earlier, the resolution of the proposed systems will depend
on many factors, including the maximum attainable frequency shift,
integration time, acoustooptic index contrast, and success in minimizing
detector noise, laser linewidth, and speckle.  In addition to resolution
improvement, there exist many possible ways to enhance the functionality and
the performance of the proposed devices.  For instance, because the phase
information is preserved, the full complex field in the object plane can be
reconstructed, potentially enabling 3D imaging via, e.g., phase-shifting
interferometry~\cite{Yamaguchi1997}.  On the performance front, the
sensitivity of the device may be improved with a subwavelength layer of
highly doped semiconductor at the front AOM facet.  When the dielectric
constant of this layer is equal to $-1$, the evanescent fields are strongly
enhanced due to resonant coupling to surface plasmons in the doped layer (a
phenomenon known as ``poor-man's superlensing''~\cite{pendry}), leading to
better SNR at the detector. Another possible way to improve the scattering
efficiency of evanescent waves is placing the sample directly in the path of
an acoustic wave, for instance, by running the wave through a microchannel
containing objects to be studied.  This approach may find many applications
in novel integrated biological/chemical detection devices.

Finally, we comment on the possibility of extending the proposed approach to
frequencies other than the mid-IR and THz bands discussed here.  While
implementing the system for lower frequencies is essentially trivial, near-IR
and optical frequencies pose a challenge.  Acoustic phonon energies in
practical devices do not approach the values necessary to produce a
substantial wave vector shift in these spectral bands.  However, the required
shift in spatial and temporal frequencies can in principle be attained by
replacing the acoustooptic medium with a nanostructured periodically moving
grating.  In this scenario, the spatial frequency shift is determined by the
periodicity of the grating while the temporal frequency shift is determined
by the speed of grating oscillation.  These parameters can be adjusted
independently to optimize performance.

\section{Conclusion}
We have proposed a system that enables detection of
sub-diffraction-limited spatial spectrum components in the far field
by utilizing scattering from an acoustic grating.  This process works
whenever the spatial frequencies of the object are comparable in scale
to the acoustic wave vector.  In its simplest implementation, the
system could aid in ``fingerprinting'' of samples based on their
subwavelength spatial features.  With the use of an additional
Bragg-shifted reference signal, it is also possible to recover the
phase of the original optical signal.  The proposed approach
has the potential to greatly enhance the specificity of mid-IR and THz
spectroscopy.

\section*{Acknowledgements}

This work was supported in part by United States Army Research Office
(USARO) Multidisciplinary University Research Initiative program awards 50342-PH-MUR and W911NF-09-1-0539.


\begin{thebibliography}{10}
\newcommand{\enquote}[1]{``#1''}

\bibitem{Testorf2010}
M.~E. Testorf and M.~A. Fiddy, \enquote{{Superresolution Imaging Revisited},}
  Advances in Imaging and Electron Physics \textbf{163}, 165--218 (2010).

\bibitem{Lukosz1967}
W.~Lukosz, \enquote{{Optical Systems with Resolving Powers Exceeding the
  Classical Limit II},} Journal of the Optical Society of America \textbf{57},
  932 (1967).

\bibitem{Kuznetsova2007}
Y.~Kuznetsova, A.~Neumann, and S.~R. Brueck, \enquote{{Imaging interferometric
  microscopy-approaching the linear systems limits of optical resolution.}}
  Optics express \textbf{15}, 6651--63 (2007).

\bibitem{Gustafsson2005}
M.~G.~L. Gustafsson, \enquote{{Nonlinear structured-illumination microscopy:
  wide-field fluorescence imaging with theoretically unlimited resolution.}}
  Proceedings of the National Academy of Sciences of the United States of
  America \textbf{102}, 13,081--6 (2005).

\bibitem{Paturzo2008}
M.~Paturzo, F.~Merola, S.~Grilli, S.~{De Nicola}, a.~Finizio, and P.~Ferraro,
  \enquote{{Super-resolution in digital holography by a two-dimensional dynamic
  phase grating},} Optics Express \textbf{16}, 17,107 (2008).

\bibitem{Durant2006}
S.~Durant, Z.~Liu, J.~M. Steele, and X.~Zhang, \enquote{{Theory of the
  transmission properties of an optical far-field superlens for imaging beyond
  the diffraction limit},} Journal of the Optical Society of America B
  \textbf{23}, 2383 (2006).

\bibitem{Dragnea2001}
B.~Dragnea, J.~Preusser, J.~M. Szarko, S.~R. Leone, and W.~D.~J. Hinsberg,
  \enquote{Pattern characterization of deep-ultraviolet photoresists by
  near-field infrared microscopy,} J. Vac. Sci. Technol. B \textbf{19},
  142--152 (2001).

\bibitem{Keilmann1999}
B.~Knoll and F.~Keilmann, \enquote{Near-field probing of vibrational absorption
  for chemical microscopy,} Nature \textbf{399}, 134--137 (1999).

\bibitem{Planken2002}
N.~C.~J. van~der Valk and P.~C.~M. Planken, \enquote{Electro-optic detection of
  subwavelength terahertz spot sizes in the near field of a metal tip,} Appl.
  Phys. Lett. \textbf{81}, 1558--1560 (2002).

\bibitem{Mendlovic1997a}
D.~Mendlovic, A.~W. Lohmann, N.~Konforti, I.~Kiryuschev, and Z.~Zalevsky,
  \enquote{{One-dimensional superresolution optical system for temporally
  restricted objects},} Applied Optics \textbf{36}, 2353 (1997).

\bibitem{Shemer1999}
A.~Shemer, D.~Mendlovic, Z.~Zalevsky, J.~Garcia, and P.~{Garcia Martinez},
  \enquote{{Superresolving Optical System with Time Multiplexing and Computer
  Decoding},} Applied Optics \textbf{38}, 7245 (1999).

\bibitem{Boyd_NLO}
R.~W. Boyd, \emph{Nonlinear Optics} (Academic Press, San Diego, 2003), 2nd ed.

\bibitem{KorpelBook}
A.~Korpel, \emph{Acoustooptics} (Marcel Dekker, New York, 1989).

\bibitem{HiFreqUltrasoundBulk}
R.~Lanz and P.~Muralt, \enquote{Bandpass filters for 8 ghz using solidly
  mounted bulk acoustic wave resonators,} IEEE Transactions on Ultrasonics,
  Ferroelectrics and Frequency Control \textbf{52}, 938 -- 948 (2005).

\bibitem{HiFreqUltrasoundSAW}
M.~B. Assouar, O.~Elmazria, P.~Kirsch, P.~Alnot, V.~Mortet, and C.~Tiusan,
  \enquote{High-frequency surface acoustic wave devices based on aln/diamond
  layered structure realized using e-beam lithography,} Journal of Applied
  Physics \textbf{101}, 114507 (2007).

\bibitem{PodolskiyNarimanovNSSL}
V.~A. Podolskiy and E.~E. Narimanov, \enquote{Near-sighted superlens,} Opt.
  Lett. \textbf{30}, 75--77 (2005).

\bibitem{Zheludev2008a}
N.~I. Zheludev, \enquote{{What diffraction limit?}} Nature materials
  \textbf{7}, 420--2 (2008).

\bibitem{OriginalGoodmanDigitalHoloPaper}
J.~W. Goodman and R.~W. Lawrence, \enquote{Digital image formation from
  electronically detected holograms,} Applied Physics Letters \textbf{11},
  77--79 (1967).

\bibitem{HoloCCDProblem1}
U.~Schnars and W.~J\"{u}ptner, \enquote{Direct recording of holograms by a ccd
  target and numerical reconstruction,} Applied Optics \textbf{33}, 179181
  (1994).

\bibitem{CucheDigitalHolo}
E.~Cuche, P.~Marquet, and C.~Depeursinge, \enquote{Simultaneous
  amplitude-contrast and quantitative phase-contrast microscopy by numerical
  reconstruction of fresnel off-axis holograms,} Appl. Opt. \textbf{38},
  6994--7001 (1999).

\bibitem{Alexandrov2006}
S.~Alexandrov, T.~Hillman, T.~Gutzler, and D.~Sampson, \enquote{{Synthetic
  Aperture Fourier Holographic Optical Microscopy},} Physical Review Letters
  \textbf{97}, 168,102 (2006).

\bibitem{Liu2002}
C.~Liu, Z.~Liu, F.~Bo, Y.~Wang, and J.~Zhu, \enquote{{Super-resolution digital
  holographic imaging method},} Applied Physics Letters \textbf{81}, 3143
  (2002).

\bibitem{LeClerc2000}
F.~{Le Clerc}, L.~Collot, and M.~Gross, \enquote{{Numerical heterodyne
  holography with two-dimensional photodetector arrays.}} Optics letters
  \textbf{25}, 716--8 (2000).

\bibitem{Yamaguchi1997}
I.~Yamaguchi and T.~Zhang, \enquote{{Phase-shifting digital holography.}}
  Optics letters \textbf{22}, 1268--70 (1997).

\bibitem{pendry}
J.~B. Pendry, \enquote{Negative refraction makes a perfect lens,}
  Phys.~Rev.~Lett. \textbf{85}, 3966--3969 (2000).

\end{thebibliography}
\end{document}